# Reducing Noise Figure and Nonlinear Penalty in Distributed Raman Amplifier System Utilizing Low-noise Forward Pumping Technique

Hiroto Kawakami, *Member, IEEE,* Kohei Saito, Akira Masuda, Shuto Yamamoto, *Member, IEEE,* and Etsushi Yamazaki, *Member, IEEE*

***Abstract*—In this paper, we experimentally and theoretically show the improvement in noise characteristics in a distributed Raman amplifier (DRA) system for wavelength division multiplexing (WDM) transmission, utilizing our proposed pumping technique. We show that forward (Fwd) pumping is clearly superior to backward (Bwd) pumping in terms of noise figure (NF) defined by amplified spontaneous emission (ASE) noise and gain. We also show that bi-directional pumping is a more desirable configuration for NF improvement. However, it is known that Bwd pumping is preferable to Fwd pumping in suppressing signal quality degradation caused by nonlinear optical effects, especially the interaction between signal and pump light. To compare these advantages and disadvantages of Fwd pumping, we conducted WDM transmission experiments using a recirculating loop including DRA and erbium doped fiber amplifiers. We measured the Q-factors of 9-channel 131.6-GBaud polarization division multiplexed probabilistically shaped 32-quadrature amplitude modulation signals while changing the ratio of Fwd pumping to Bwd pumping in the DRA. By introducing the previously proposed low-noise Fwd pumping technique, a higher Q-factor could be achieved with a lower signal launch power, even when the total Raman gain remained constant. A Q-factor improvement of 0.4 dB and signal launch power reduction of more than 3.4 dB were simultaneously achieved. We also show that when Fwd pumping was performed with a conventional pump light source, the disadvantage of Fwd pumping was more noticeable than its advantage.**

## I. INTRODUCTION

DISTRIBUTED Raman amplifiers (DRAs) are very attractive for long-haul and/or wideband optical transmission systems because their distributed gain can reduce transmission loss and allow for flexible control of the center wavelength of the gain spectrum. However, the gain of DRAs is limited by the maximum pump power that can be launched to the gain medium, i.e., the transmission line. To mutually compensate for the shortcomings, hybrids of DRAs and rare-earth-doped fiber amplifiers, especially erbium-doped fiber amplifiers (EDFAs), have been widely investigated [1], [2], [3], [4], [5], [6], [7]. As is well known, to pump the DRA, forward (Fwd) pumping, backward (Bwd) pumping, and bi-directional pumping can be used [8], [9], [10], [11], [12]. In terms of noise figure (NF) defined by amplified spontaneous emission (ASE) noise and gain, Fwd pumping is superior to Bwd pumping [10], [11]. Note that, in [10] and [11], the NF in DRAs was defined differently. To evaluate the NF in a hybrid system of a DRA and EDFA, it is required to use a properly defined NF. Another consideration is that it is known that a DRA using Fwd pumping is prone to signal degradation unrelated to ASE. First, because the maximum intensity of the optical signal(s) around the input end of the DRA tends to be higher than when using only Bwd pumping [8], [9], [12], the optical signal(s) tends to induce nonlinear optical effects. Second, since the pump light and signal light propagate in the same direction, interaction between them tends to degrade signal quality [8], [13], [14], [15], [16]. Here, because Raman gain depends on the polarization of the signal and pump light [17], [18], [19], the pump light should be depolarized by two polarization-combined pump light sources [14], [15] or a depolarizer utilizing optical delay [20], [21], [22], [23]. We previously reported that polarization-multiplexed pump light can generate unstable "synthesized-polarization" that can induce instability in the Raman gain [22], [23], [24]. We previously proposed a low-noise pumping technique with two polarization-combined pump light sources that can suppress signal degradation induced by this gain instability [24]. In more recent work, we experimentally showed that a DRA with bi-directional pumping that uses our proposed low-noise Fwd pumping and conventional Bwd pumping simultaneously is

This paragraph of the first footnote will contain the date on which you submitted your paper for review, which is populated by IEEE. *(Corresponding author: Hiroto Kawakami.).*

Hiroto kawakami is with NTT Network Innovation Laboratories, NTT, Yokosuka 2390847, Japan (e-mail: hiroto.kawakami@ntt.com).

Kohei Saito is with NTT Network Innovation Laboratories, NTT, Yokosuka 2390847, Japan (e-mail: kohei.saito@ntt.com).

Akira Masuda is with NTT Network Innovation Laboratories, NTT, Yokosuka 2390847, Japan (e-mail: ak.masuda@ntt.com).

Shuto Yamamoto is with NTT Network Innovation Laboratories, NTT, Yokosuka 2390847, Japan (e-mail: shuto.yamamoto@ntt.com).

Etsushi Yamazaki is with NTT Network Innovation Laboratories, NTT, Yokosuka 2390847, Japan (e-mail: etsushi.yamazaki@ntt.com).

TABLE I
COMPARISON OF FWD, BWD AND BI-DIRECTIONAL PUMPING TECHNOLOGIES

| Ref. | Evaluation method | Pump light source for Fwd pumping | Evaluation of NF or OSNR | | Evaluation of signal quality after demodulation | | |
|---|---|---|---|---|---|---|---|
| | | | DRA only | Hybrid system of DRA and EDFA | Optical channel, signal format (*) | Single span | Multi-span |
| [11] | Simulation | LD | Fwd was better | N. E. (**) | N. E. | N. E. | N. E. |
| [10] | Mainly simulation | Conventional LD | Fwd was better | N. E. | Binary signal | Fwd was preferable | N. E. |
| [4] | Experiment | Conventional LD | N. E. | N. E. | WDM 120 Gb/s QPSK | Bwd only was better | |
| | | Broadband light utilizing 10 km fiber | | | | Bi-directional was better | |
| [7] | Experiment | Semiconductor incoherent pump | N. E. | N. E. | WDM 100 Gbaud 16-QAM | Bi-directional was better | |
| [8] | Experiment | ASE of EDFA | Fwd was better | N. E. | Single channel 9.2 Gbaud BPSK | Bi-directional was better | N. E. |
| [25] | Experiment | Our low-noise pump unit (1470 nm) | N. E. | Bi-directional was better | Single channel 32 Gbaud 16-QAM | Bi-directional was better | N. E. |
| This work | Simulation & Experiment | Our low-noise pump unit (1430+1450 nm) | Bi-directional is better | | WDM, 131.6 Gbaud PS 32-QAM | Bi-directional is better | |

(*) All measured signals were polarization division multiplexed, (**) N.E. : Not evaluated

effective in improving the Q-factor [25].

However, in [25], we set several conditions in the experimental setup to demonstrate the effects of DRA more clearly. First, to obtain a high Raman gain and avoid saturation of the Q-factor due to electrical circuit noise (see "result and discussion" in [25]), a 125-km dispersion shifted fiber (DSF) was used for the DRA. Next, in [25], a single wavelength signal light was used instead of a wavelength division multiplexed (WDM) signal to investigate only the interaction between the pump and the signal light, rather than the interaction between signals of adjacent wavelength.

In this paper, to investigate our proposed low-noise pumping technique under more general conditions, we conducted WDM transmission experiments using a recirculating loop including a DRA and EDFAs. In these experiments, our low-noise pump unit proposed in [24] was improved for WDM transmission systems. The gain medium (transmission line) of the DRA was 100.6-km G.654.E, low-loss and large-core cut-off shifted fiber widely investigated for WDM transmission systems [1]. First, we show the noise characteristics of this DRA alone and the noise characteristics of a hybrid system of this DRA and EDFAs. Improvements in the optical signal to noise ratio (OSNR) due to Fwd pumping are calculated using a properly defined NF. Finally, we show the measured Q-factors of the transmitted 9-channel 131.6-GBaud polarization division multiplexed (PDM) probabilistically shaped (PS) 32-quadrature amplitude modulation (QAM) signals, while changing the ratio of Fwd pump power and Bwd pump power in the DRA. The experimental results show that the advantages of Fwd pumping become more significant than its disadvantages when using the proposed low-noise Fwd pumping technique, though the disadvantages of Fwd pumping become more significant when using the conventional pump light source. Table 1 is a brief comparison of Fwd, Bwd, and bi-directional pumping reported in the past literature and this study. It shows that our proposed Fwd-pumping technique has an advantage even for high-order PS QAM signals transmitted over multi-span lines.

## II. CONSTRUCTION OF DRA WITH OUR PROPOSED PUMPING TECHNIQUE

Fig. 1 shows the construction of the DRA used in this study. 100.6-km single mode fiber (ITU-T G. 654. E) was used for the gain medium. Two WDM couplers (WDMcs) were used to multiplex and demultiplex the pump light and signal light. In this paper, "intrinsic loss of the DRA" is defined as the ratio of the optical power measured at reference points Ra and Rd shown in Fig. 1 with the pump light turned off. The denominator is the optical power at Ra; therefore, intrinsic loss takes a value between 0 and 1. In this study, the intrinsic loss of the DRA shown in Fig. 1 was 0.017 = -17.8 dB. With the pump light turned on, the "net gain of the DRA" is defined using the same calculation. Though the on/off gain is always greater than 1 (0 dB), the net gain can be less than 1 because of the intrinsic loss. In fact, in this study, it takes a value smaller than 1.

The pump light for Fwd pumping and Bwd pumping was generated by our proposed Fwd-pump unit and conventional Bwd-pump unit, respectively. In this study, both of these pump units had four multimode pump laser diodes (LDs), denoted as 1X, 1Y, 2X and 2Y. The output light from these four LDs was polarization multiplexed using two polarization beam combiners (PBCs) and wavelength division multiplexed using two WDMcs within each pump unit. In our previous works [24], [25], each pump unit had only two pump LDs due to equipment limitations, so wavelength division multiplexing of the pump light was not used. In this study, we were able to achieve a wide gain band using pump lights with different center wavelengths, around 1430 nm and around 1445 nm. The four pump LDs in the Bwd-pump unit were conventional pump LDs containing fiber Bragg gratings (FBGs) [14], [15]. In comparison, the four pump LDs in our proposed Fwd-pump unit had no FBGs [22], [23], [24], [25]. The currents and temperatures of these eight pump LDs were adjusted using a controller to reach the target conditions described later. Control of the pump power and adjustment of the Raman gain were performed by polarization-maintaining variable optical attenuators (VOAs), without changing the LD currents [24], [25]. Fig. 2 (a) shows the measured optical spectra of these four multimode pump LDs in the proposed Fwd pump unit. Because these LDs had no FBG, the wavelengths of each longitudinal mode and the center wavelength of their envelopes can be controlled by the temperature and/or current of the LD. We set the wavelengths so that the overlap of the envelopes of LD_1X and LD_2X (similarly, LD_1Y and LD_2Y) was as small as possible, and the longitudinal modes of LD_1X and LD1_Y (similarly, LD_2X and LD_2Y) were interleaved. This wavelength arrangement can reduce noise transfer from polarization multiplexed pump light to signal light [24]. Fig. 2 (b) shows the measured optical spectra of the four multimode pump LDs in the conventional Bwd pump unit. Because these LDs had FBGs, the envelopes were narrow, and their wavelengths were fixed, though each wavelength of the longitudinal mode can be controlled by temperature and/or current. Though we also set the wavelengths of the longitudinal modes of LD_1X and LD_1Y (similarly, LD_2X and LD_2Y) to be interleaved, partial overlap of the spectral cannot be avoided because of the fine structure generated by the FBG [15]. Though these partial overlaps do not cause major problems with Bwd pumping, they cause non-negligible degradation in the signal quality with Fwd pumping [24], [25] as will be shown in later.

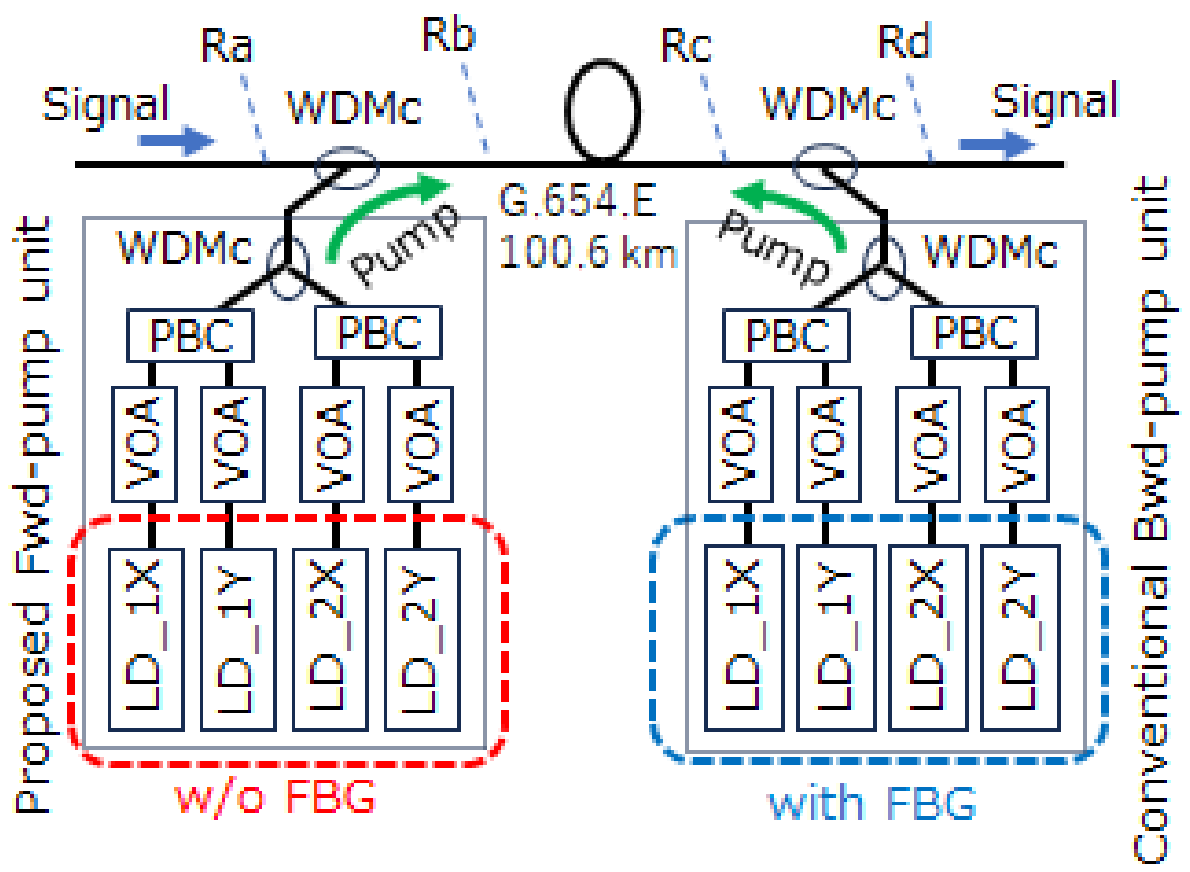

Fig. 1. Construction of DRA used in this study.

## III. Noise Induced by Interaction between Signal and Pump Light

In this section, we show the measured relative intensity noise (RIN) of probe light amplified using the DRA. The configuration of the DRA used for this measurement was almost the same as that shown in Fig. 1. However, only in this section, the length of the G. 654. E fiber was 50.3-km instead of 100.6-km. This was because, to ensure the measurement accuracy of the RIN measurement system, it was necessary to reduce the intrinsic loss of the DRA. The probe light used for the measurement was single-polarization, continuous wave (CW) light with a wavelength of 1535 nm. The power of the probe light was 0.1 dBm at reference point Rb in Fig. 1. In this section, the loss of the VOAs shown in Fig. 1 was adjusted so that the on/off gain of the DRA was 6.7 dB in all experiments. The measured RIN spectrum of the probe light before launching to the DRA is shown in Fig 3(a). Above 10 GHz, RIN was approximately -150 dB/Hz, and below 1GHz, it was below the measurement limit. The measured RIN spectrum after amplification with the DRA is shown in Figs. 3(b) and (c). In Fig. 3(b), only the proposed Fwd pump unit was used. In Fig. 3(c), only the conventional Bwd pump unit was used. These RIN spectra were almost identical. Although the RIN after the amplifier peaked at 8 GHz and increased at a lower frequency compared with Fig. 3(a), the measured RIN was less than -140 dB/Hz. For comparison, we also conducted an experiment using a DRA in which the proposed pump unit and conventional pump unit shown in Fig. 1 were replaced with each other. The measured RIN spectra after amplification are shown in Figs. 4(a), (b), and (c). In Fig. 4(a), only Fwd pumping was used with the conventional pump unit. In Fig. 4(b), only the Bwd pumping was used with proposed pump unit. When only Bwd pumping was used, there was no significant difference in the RIN spectrum of the probe light whether the proposed pump unit (Fig. 4(b)) or the conventional pump unit (Fig. 3(c)) was used. Furthermore, when using the proposed pump unit, there was no significant difference in the RIN spectrum of the probe light whether Fwd pumping (Fig.3 (b)) or Bwd pumping (Fig.4 (b)) was performed. In contrast, when Fwd pumping was performed with the conventional pump unit, a highly irregular RIN was superimposed on the low frequencies (Fig.4 (a)). As mentioned in the previous section, we set the wavelengths of the longitudinal modes of X and Y polarizations to be interleaved, though they were partially overlapped in the conventional pump unit. For comparison, we also conducted an experiment in which the longitudinal modes of a conventional pump unit were completely overlapped and used for Fwd pumping. Fig. 4(c) shows the result. The RIN spectrum of the probe light was significantly degraded [14], [15]. This was because the overlapping longitudinal modes induced random fluctuations in the Raman gain with frequency components ranging from DC to several GHz and above [24]. Figs 3 and 4 newly confirmed that our proposed technique was effective at suppressing this noise even when the pump light was wavelength division multiplexed. Hereinafter, unless otherwise specified, the configuration of the DRA used in the experiment was the same as that shown in Fig. 1, and the gain medium was 100.6-km.

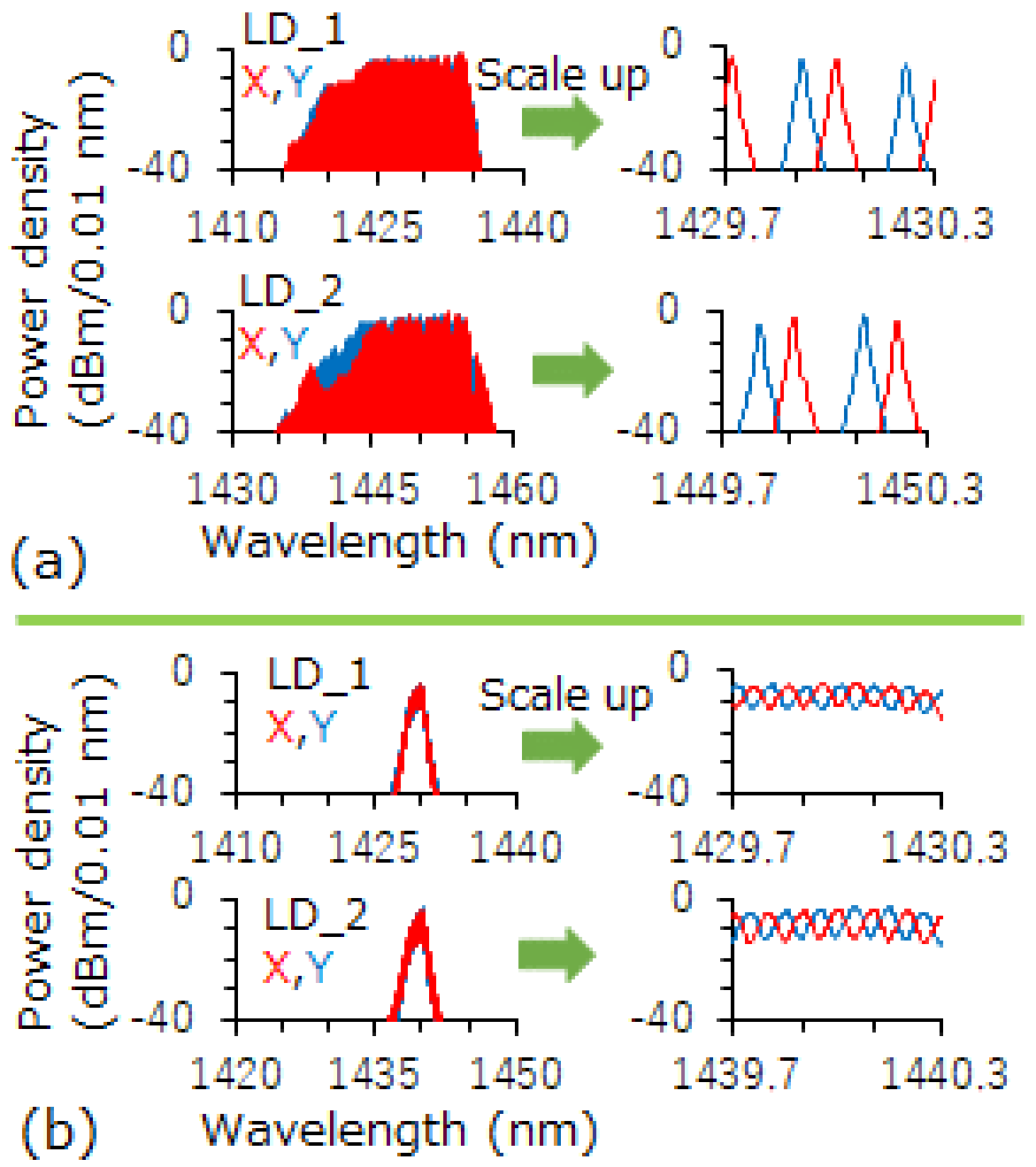

Fig. 2. Measured optical spectra of eight pump LDs. (a): Four pump LDs without FBGs used in proposed Fwd pump unit. (b): Four conventional pump LDs containing FBGs used in conventional Bwd pump unit.

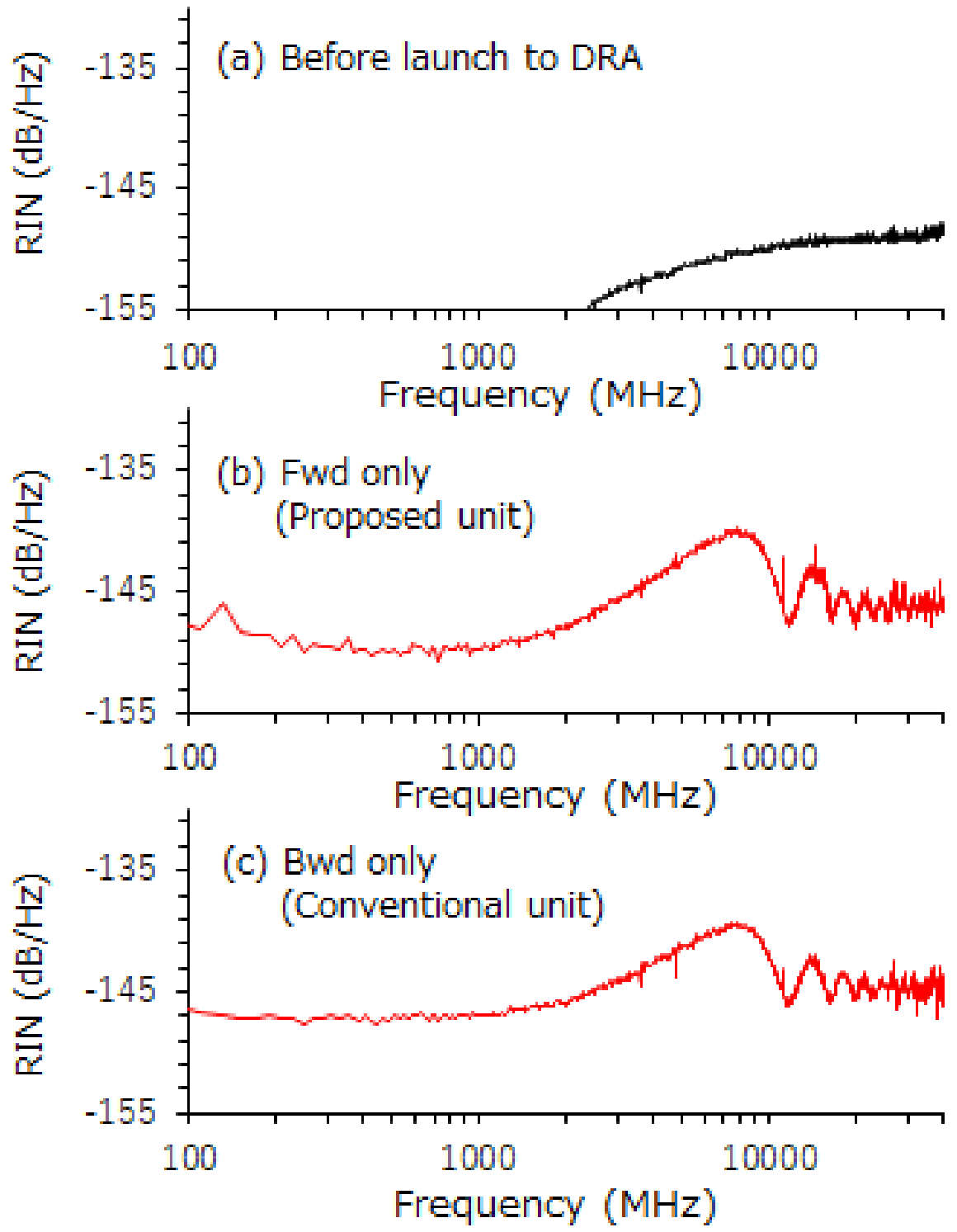

Fig. 3. Measured RIN of probe light before and after launching to DRA. For this measurement, 50.3-km fiber was used instead of 100.6-km fiber in Fig. 1. (a): before launch to DRA, (b): after amplification with DRA using only Fwd pumping with proposed pump unit, (c): after amplification with DRA using only Bwd pumping with conventional pump unit.

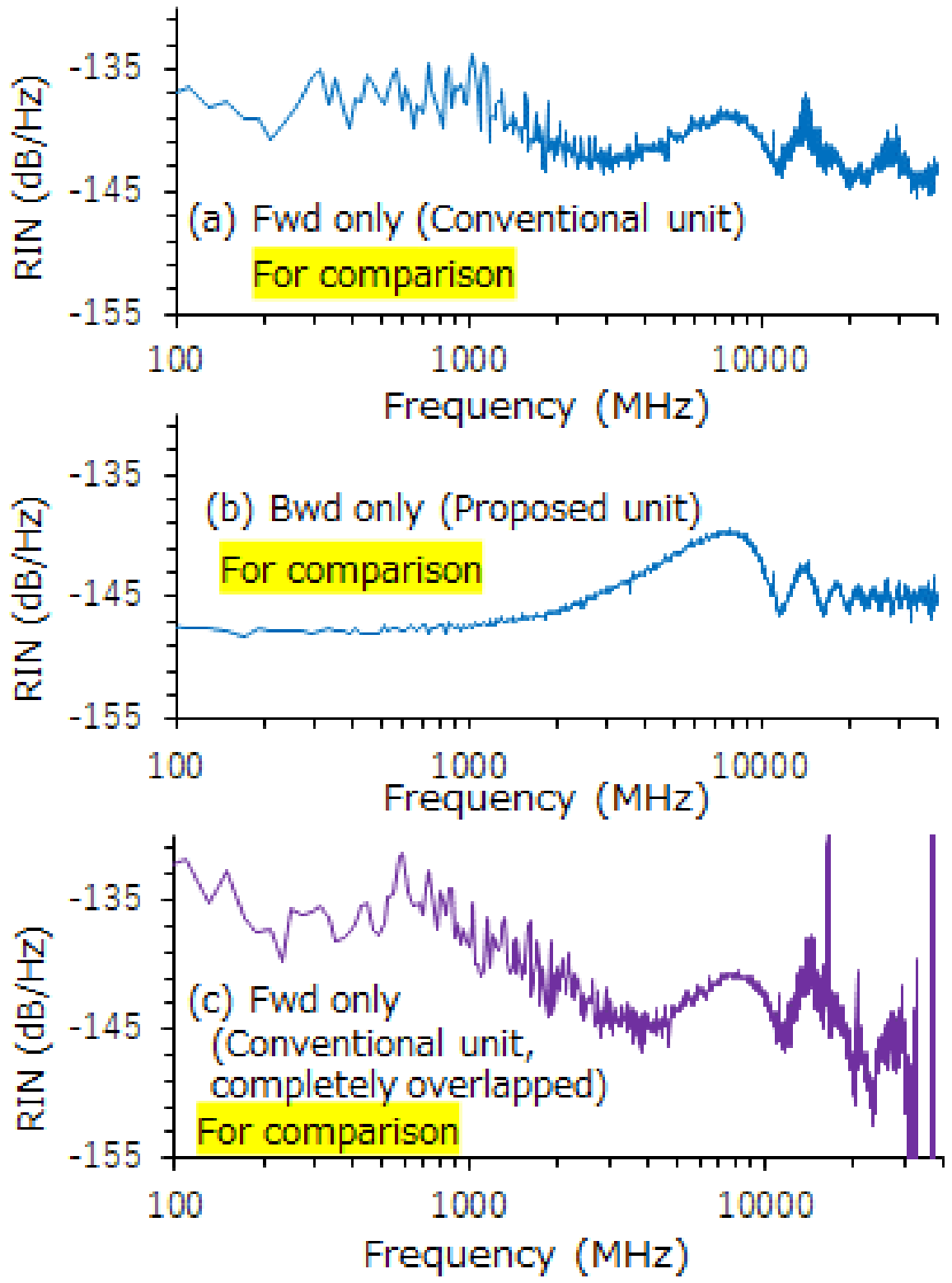

Fig. 4. Measured RIN of probe light after amplification with DRA. In this measurement, 50.3-km fiber was used, and arrangement of pump units was changed from the configuration shown in Fig 1 for comparison. (a): using only Fwd pumping with conventional pump unit, (b): using only Bwd pumping with proposed pump unit, (c): using only Fwd pumping with conventional pump unit modified so that each longitudinal mode completely overlapped.

## IV. Measured Gain and noise figure of DRA

In this section, to compare Fwd pumping and Bwd pumping, we show the measured gain and NF of the DRA shown in Fig. 1 using three different pumping methods: Fwd pumping only, Bwd pumping only, and bi-directional pumping. The pump power for Fwd pumping measured at the reference point Rb (see Fig. 1) was zero or 391 mW. The pump power for Bwd pumping measured at reference point Rc was zero or 340 mW. To facilitate comparison, these pump powers were adjusted by the VOAs so that the small signal gains (unsaturated gains) for Fwd pumping only and Bwd pumping only were almost identical. The wavelength of the probe light was 1535 nm, and the signal launch powers were -6.1 dBm to +4.7 dBm. In this paper, “signal launch power” is defined as the optical power at the reference point Ra. Fig. 5 shows the measured on/off gain. When the signal launch power was under +5 dBm, the measured on/off gains were 6.4 dB (Fwd pumping only), 6.3 dB (Bwd pumping only), and 12.7 dB (bi-directional pumping). The simple relationship (6.4+6.3) [dB] = 12.7 [dB] shows that the gain caused by Fwd pumping and gain caused by Bwd pumping can be treated as independent parameters with no interaction. The net gain of the DRA can be calculated by the intrinsic loss (-17.8 dB) and on/off gain. For example, when the on/off gain was 6.4 dB, the net gain of this DRA was -11.4 dB, lower than 1.

Fig. 6 shows the noise spectra generated by the ASE of the DRA, measured at reference point Rd. The powers of the pump lights were the same as those used in the measurement in Fig. 5. When performing the measurements shown in Fig. 6, the probe light was turned off to make the ASE noise easier to see. When comparing the case of Fwd pumping only and Bwd pumping only, the shape of the spectrum of the ASE noise was different because the spectra of the Fwd pump light and Bwd pump light were not identical (see Fig. 2). However, all spectra had a maximum at approximately 1535 nm (indicated by the dashed line), which was the probe wavelength used in the RIN measurement and on/off gain measurement. Fig. 6 shows that highest gain can be achieved around this wavelength. The ASE noise measured using only Fwd pumping was drastically lower compared with using only Bwd pumping, although their small signal gains were almost identical (see Fig. 5). This can be explained as follows. When only Fwd pumping was used, ASE noise was generated mainly near the input end of the DRA and attenuated near the output end. In comparison, when only Bwd pumping was used, ASE noise was generated mainly near the output end of the DRA without attenuation. The ASE noise measured using bi-directional pumping was of similar shape and power as when only Bwd pumping was used, although the gain using bi-directional pumping was 6.4 dB higher than the gain using only Bwd pumping. These results suggest that NF can be improved by introducing the Fwd pumping.

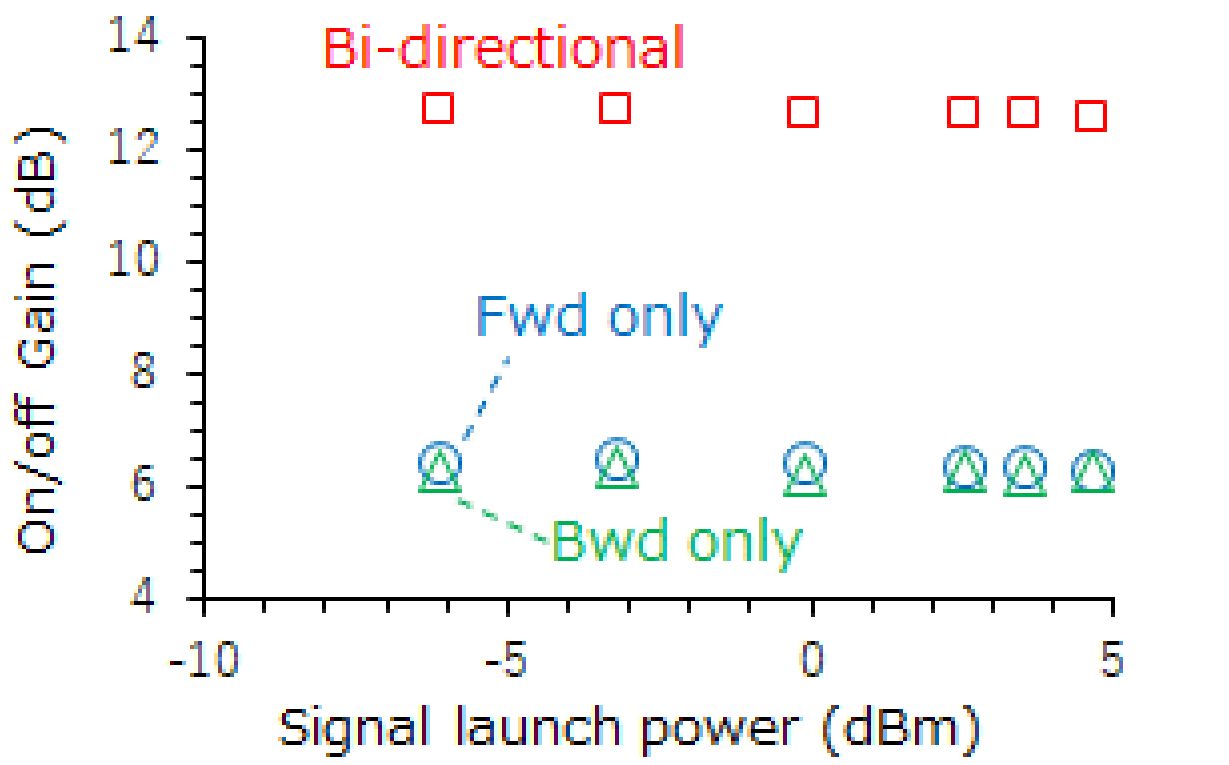

Fig. 5. Measured on/off gain of the DRA at 1535 nm as the function of the signal launch power. Circle: Fwd pumping (391 mW) only, triangle: Bwd pumping (340 mW) only, square: bi-directional pumping.

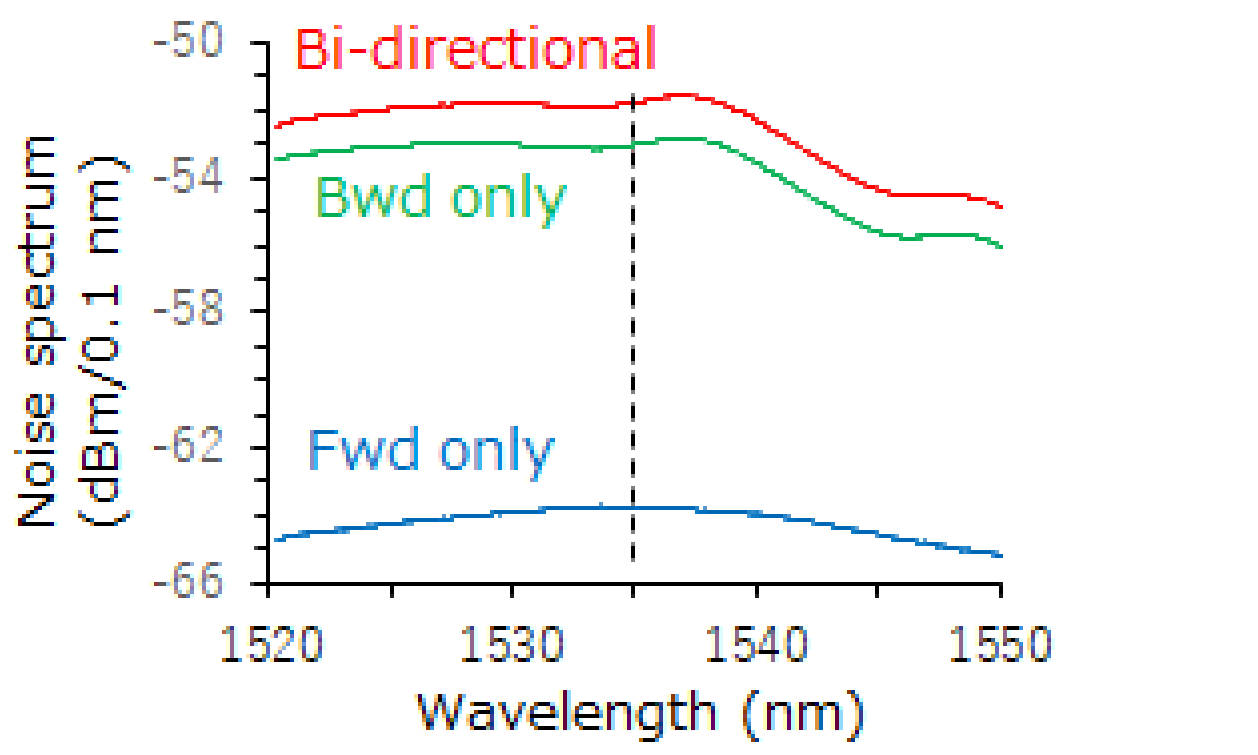

Fig. 6. Measured ASE noise spectra at Rd generated by the DRA. Blue: Fwd pumping (391 mW) only, green: Bwd pumping (340 mW) only, red: bidirectional pumping.

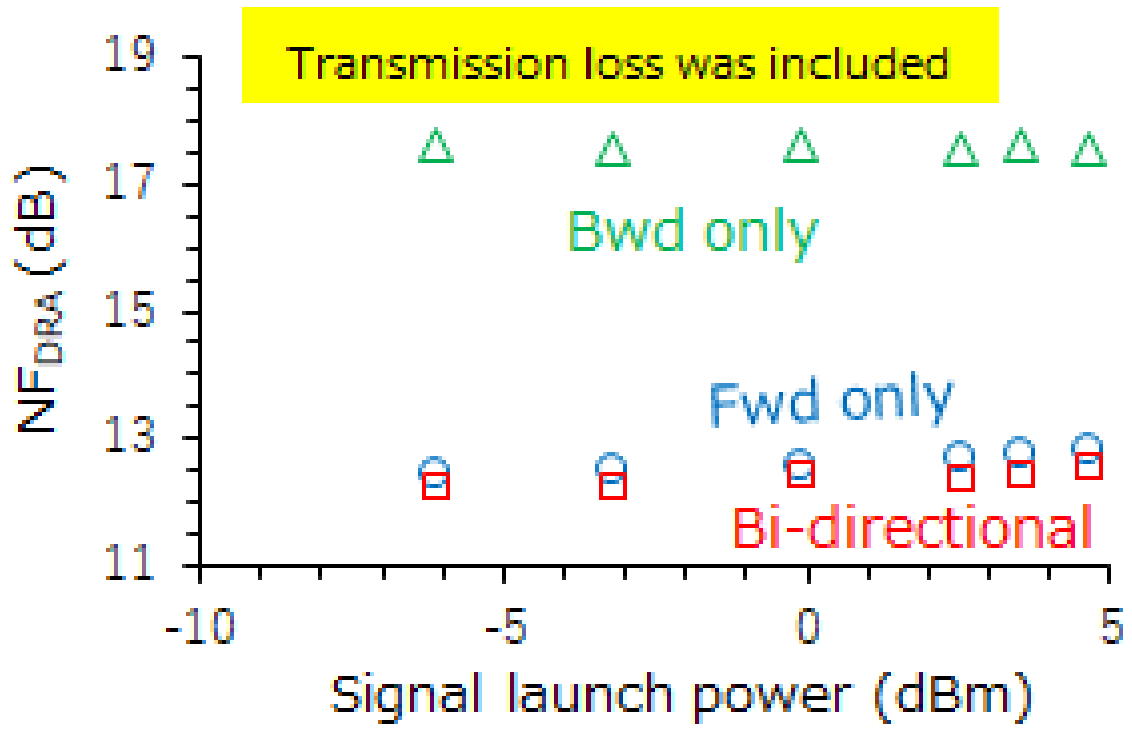

Fig. 7. Calculated $NF_{DRA}$ using measured $ASE_{DRA}$ and $G_{DRA}$ of the DRA shown in Fig. 1. Circle: Fwd pumping (391 mW) only, triangle: Bwd pumping (340 mW) only, square: bi-directional pumping.

Before calculating the NF, we define the "NF of the DRA" used in this paper. There are several definitions for the NF of a DRA. One definition calculates the NF using the on/off gain [11]. It is denoted as "equivalent NF" in [26]. This definition is used for evaluating an amplifier placed virtually at the output end of the DRA, and this virtual amplifier does not have the intrinsic loss that a real DRA has. Another definition considers the intrinsic loss of the DRA and calculates the NF using the net gain [10]. It is denoted as "intrinsic NF" in [26]. This definition can treat the entire DRA (from Ra to Rd in Fig. 1) as one giant optical amplifier. In this paper, the NF of the DRA means "intrinsic NF" and is calculated as follows. (Note that, in this paper, the variables used in all equations with numbering are linear, not dB.)

$$NF_{DRA} = ASE_{DRA}/(h \cdot \upsilon \cdot \Delta\upsilon \cdot G_{DRA}) + 1/G_{DRA} \quad (1)$$

where $ASE_{DRA}$ is the power density of the ASE noise summed over all polarization states, $h\upsilon$ is the photon energy of the signal light, $\Delta\upsilon$ is the optical bandwidth of the $ASE_{DRA}$, and $G_{DRA}$ is the net gain of the DRA [10]. Because net gain is the product of on/off gain and intrinsic loss,

$$NF_{DRA} = NF_{eq}/Loss_{int} > NF_{eq} \quad (2)$$

where $NF_{eq}$ is an equivalent NF, and $Loss_{int}$ is the intrinsic loss of the DRA ($0 < Loss_{int} < 1$). If the intrinsic loss of the DRA is 0.01 = -20 dB, $NF_{DRA}$ is 20 dB larger than the equivalent NF (see Fig. 5 in [26]). When the length of the gain medium of the DRA is very short, $ASE_{DRA}$ is close to zero, and $G_{DRA}$ is close to 1 (0 dB), so then $NF_{DRA}$ is close to 1 (0 dB). This is lower than the quantum limit of NF, which is 2 (3 dB), but this is not a contradiction because the quantum limit of NF is a value assuming a large gain [26], [27], [28].

Note that the second term in (1) does not depend on the ASE power. This term represents shot noise. Applying (1), it is possible to define the noise figure of a passive optical device that has insertion loss and no gain (for example, a VOA, unexcited long optical fiber, optical 3-dB coupler, etc.). Since these do not generate ASE, the NF of these optical devices can be defined as follows.

$$NF_{LOSS} \equiv 1/G_{LOSS} > 1 \quad \text{(without ASE noise)} \quad (3)$$

where $G_{LOSS}$ is the (output power)/(input power) of the optical device. The validity of this definition will be shown in (6).

Fig. 7 shows the calculated $NF_{DRA}$ of the DRA shown in Fig. 1, substituting the measured $G_{DRA}$ and $ASE_{DRA}$ into (1). The powers of the pump lights were the same as those used in the measurement in Figs. 5, 6. The measured $NF_{DRA}$ with the small signal launch power was 17.4 = 12.4 dB when only Fwd pumping was used, 57.5 = 17.6 dB when only Bwd pumping was used, and 17.0 = 12.3 dB when bi-directional pumping was used. When using bi-directional pumping, Fwd pumping acted as a pre-amplifier, and the pump light reached almost the entire 100.6-km fiber, resulting in the best $NF_{DRA}$ and $G_{DRA}$. Note that these $NF_{DRA}$ were higher than the typical NF of an EDFA because they reflect the transmission loss of 100.6-km optical fiber.

## V. NF of Amplifier Chain Consisting of DRA and EDFA

As mentioned in Section I, it is difficult to configure a multi-span optical transmission system using only a DRA, so transmission spans consisting of a hybrid of a DRA and EDFA(s) are often chained together. In this section, we calculate the NF of such a multi-span transmission line. Moreover, we estimate the OSNR of the transmitted signal. Fig. 8 shows the configuration model of the N-span transmission system assumed in the simulation. In this figure, LS means a CW light source, L1-L5 mean an optical device(s) that has insertion loss(es) and no gain. "L1 and EDFA(s)" means the EDFA is connected before and/or after L1 to compensate for the optical loss(es) of L1. $R_i$ ("i" is an integer from 0 to 7) is a reference point for explanation. In this paper, the total gain and total NF of all optical components existing between reference points $R_i$ and $R_j$ are defined as $G_{ij}$ and $NF_{ij}$, respectively.

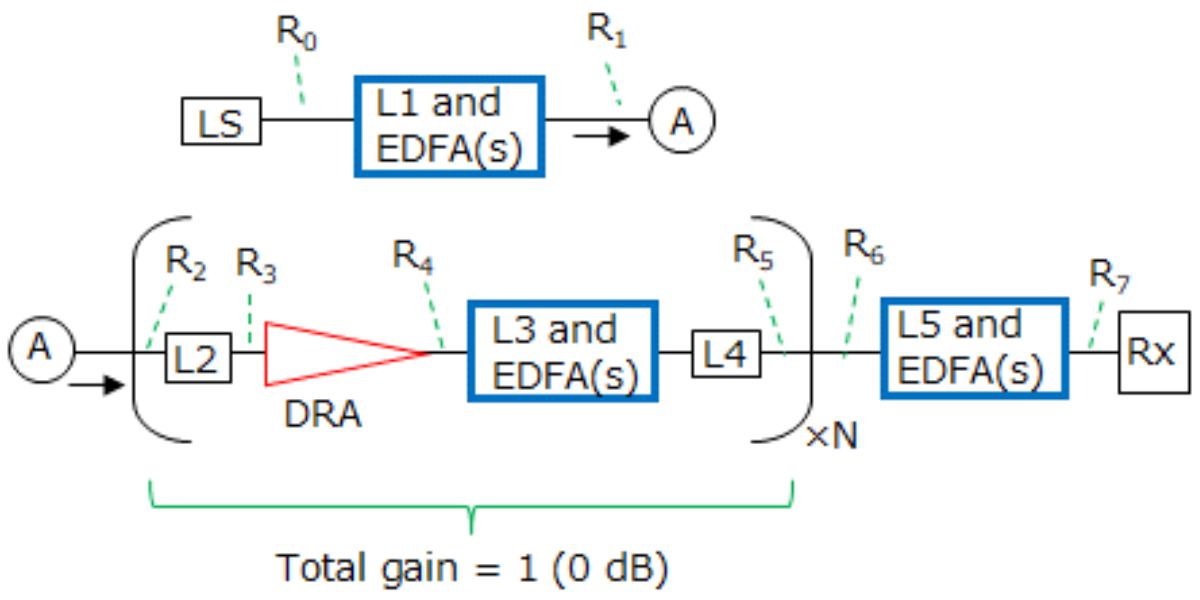

Fig. 8. Configuration model of N-span transmission system assumed in simulation. LS: light source, L1-L5: device with insertion loss.

Our first target is to describe $NF_{07}$ as a function of the signal launch power, $G_{DRA}$, $NF_{DRA}$, and the number of spans $N$. In the previous section, we defined the signal launch power as the optical power at reference point Ra in Fig. 1. According to the notation of Fig. 8, the signal launch power is the optical power at reference point $R_3$. In a real optical transmission system, L1 corresponds to an optical modulator with insertion loss and modulation loss. When a WDM signal is used, L1 also includes a WDM coupler or arrayed waveguide grating. L2 is a VOA used to control the signal launch power at $R_3$. In the simulation, an optical switch and couplers were also included in L2 because these optical devices were used in the recirculating loop for the verification experiment. Because L2 has insertion losses and no ASE noise, using (3), $G_{23}$ and $NF_{23}$ can be described as

$$NF_{23} = 1/G_{23} > 1\ . \tag{4}$$

$G_{34}$ and $NF_{34}$ are the same as the $G_{DRA}$ and $NF_{DRA}$ shown in the previous section, respectively. Using the well-known NF addition formula [11], $NF_{24}$ can be described as

$$NF_{24} = NF_{23} + (NF_{34} - 1)/G_{23}\ . \tag{5}$$

Using (4) - (5), $NF_{24}$ can be described as

$$NF_{24} = NF_{34}/G_{23} > NF_{34}\ . \tag{6}$$

Equation (6) corresponds to the relationship between equations (2.118) and (2.153) of [28]. This result shows that (3) is valid. L3 corresponds to a gain equalizer (GEQ) that suppress a gain tilt. In the simulation, a polarization scrambler that cancels polarization dependent loss in the recirculating loop was also included in L3. L4 corresponds to the VOA used to adjust $G_{25}$ = 1 (0 dB). L5 corresponds to an optical bandpass filter or wavelength selective switch when a WDM signal is transmitted. In the simulation, L5 also includes an optical coupler that guides signal light from the recirculating loop to the outside. By repeating similar calculations, $G_{07}$ and $NF_{07}$ can be described as

$$G_{07} = G_{01} \cdot G_{67} \tag{7}$$

$$NF_{07} = NF_{01} + N \cdot (NF_{25} - 1)/G_{01} + (NF_{67} - 1)/G_{01} \tag{8}$$

$$NF_{25} = \{NF_{DRA} + (NF_{45} - 1)/G_{DRA}\}/G_{23}. \tag{9}$$

Because $G_{25}$ = 1, $G_{07}$ is independent from $G_{DRA}$. As mentioned above, $G_{DRA}$ depends on both the pump power and the intrinsic loss. When the transmission length is short and the intrinsic loss is small, $G_{DRA}$ becomes larger even if the on/off gain of the DRA is small. In this case, as shown in (9), the influence of $NF_{DRA}$ on $NF_{25}$ increases, so the optimizing $NF_{DRA}$ by pumping direction becomes more important. In contrast, when the transmission length is long and the intrinsic loss is large, $G_{DRA}$ becomes smaller even if the on/off gain of the DRA is large, so $NF_{45}$ becomes more dominant than $NF_{DRA}$, though the on/off gain of the DRA is still useful to reduce $NF_{25}$ and $NF_{07}$ [25].

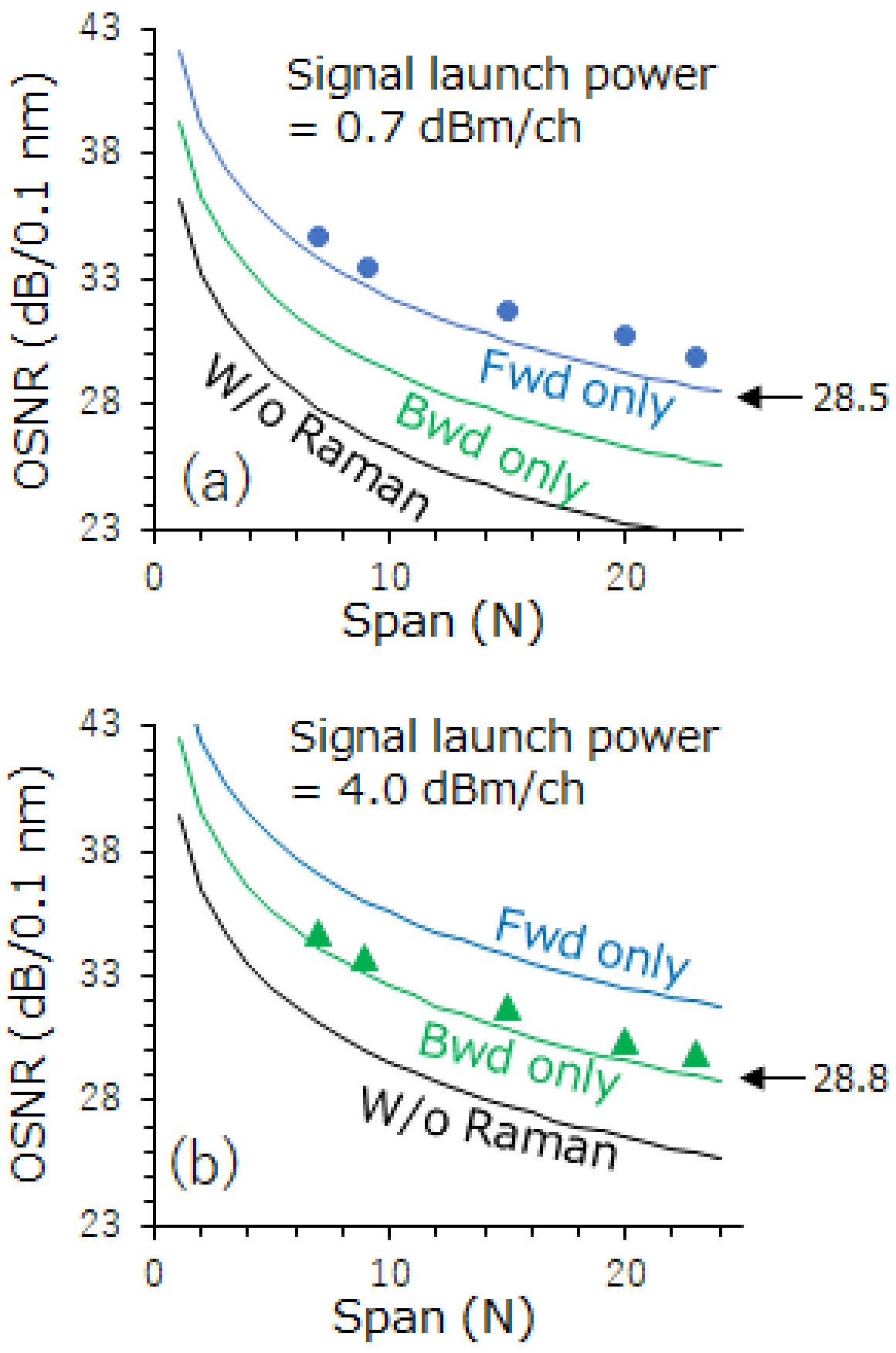

Fig. 9. Simulated (solid line) and measured (symbol) $OSNR_7$ of the 9-ch WDM signal. Construction and gain of DRA are same as in Figs. 1-3. (a): signal launch power is 0.7 dBm/ch. (b): signal launch power is 4.0 dBm/ch. Blue: Fwd pumping only, green: Bwd pumping only, black: without pumping.

Our next target is to describe the OSNR after transmission at $R_7$. Equation (1) can be rewritten as

$$NF_{07} = ASE_7/(h \cdot \upsilon \cdot \Delta\upsilon \cdot G_{07}) + 1/G_{07} \tag{10}$$

where $ASE_7$ is the power density of the ASE noise measured at $R_7$. The OSNR at $R_7$ can be calculated from (11) and (12).

$$OSNR_7 \equiv Signal_7/ASE_7 \tag{11}$$

$$Signal_7 = Signal_0 \cdot G_{07} \tag{12}$$

where $OSNR_i$ and $Signal_i$ are the OSNR and signal power measured at reference point $R_i$, respectively. Using these equations, $OSNR_7$ can be described as

$$OSNR_7 = Signal_0/\{h \cdot \upsilon \cdot \Delta\upsilon(NF_{07} - 1/G_{07})\}. \tag{13}$$

Using (13), $OSNR_7$ is simulated as a function of the number of spans N, signal launch power at $R_3$, and pumping method. Unfortunately, the characteristics of the EDFA in the recirculation loop, especially the gain dynamics between the (n)-th and (n+1)-th round, were complex, making it difficult to accurately predict their impact on gain and NF. For this reason, estimated values were substituted for some parameters such as $NF_{45}$. In the simulation, we assumed that $G_{DRA}$ = 0.073 = -11.4 dB (the on/off gain of the DRA is 6.4 dB) when pump light was Fwd only, $G_{DRA}$ = 0.071 = -11.5 dB (the on/off gain of the DRA is 6.3 dB) when pump light was Bwd only. These gains assumed in the simulation are the same as the measured gain shown in Fig. 5. In the simulation, the transmitted signal is assumed to be a 9-channel WDM signal. This number of wavelengths was matched to the number of wavelengths used in the experimental setup used for verification. The solid lines in Fig. 9 show the results of the simulation. The simulated $OSNR_7$ using only Fwd pumping (blue) was approximately 3 dB higher than using only Bwd pumping (green). When the signal launched power was 0.7 dBm/ch (Fig. 9 (a)), $OSNR_7$ using Fwd pumping only was 28.5 dB/0.1 nm at $N$ = 24. When the signal launched power was 4.0 dBm/ch (Fig. 9 (b)), $OSNR_7$ using Bwd pumping only was 28.8 dB/0.1 nm at $N$ = 24. Although the signal launched power using Fwd pumping only was 3.3 dB lower than Bwd pumping only, almost identical OSNRs were obtained. For comparison, the simulated $OSNR_7$ without Raman pumping are also shown (black). In Fig. 9, the OSNR measured experimentally using the recirculating loop is also shown by symbols. The circle and triangle indicate Fwd pumping only and Bwd pumping only, respectively. The gain of the DRA used in the experiment was the same as in the simulation. The measured OSNR almost agreed with the simulations. These results show that a higher OSNR could be achieved with a lower signal launched power by using Fwd pumping in the 100.6-km DRA, despite the condition that the on/off gain was about 6.4 dB for both Fwd pumping and Bwd pumping.

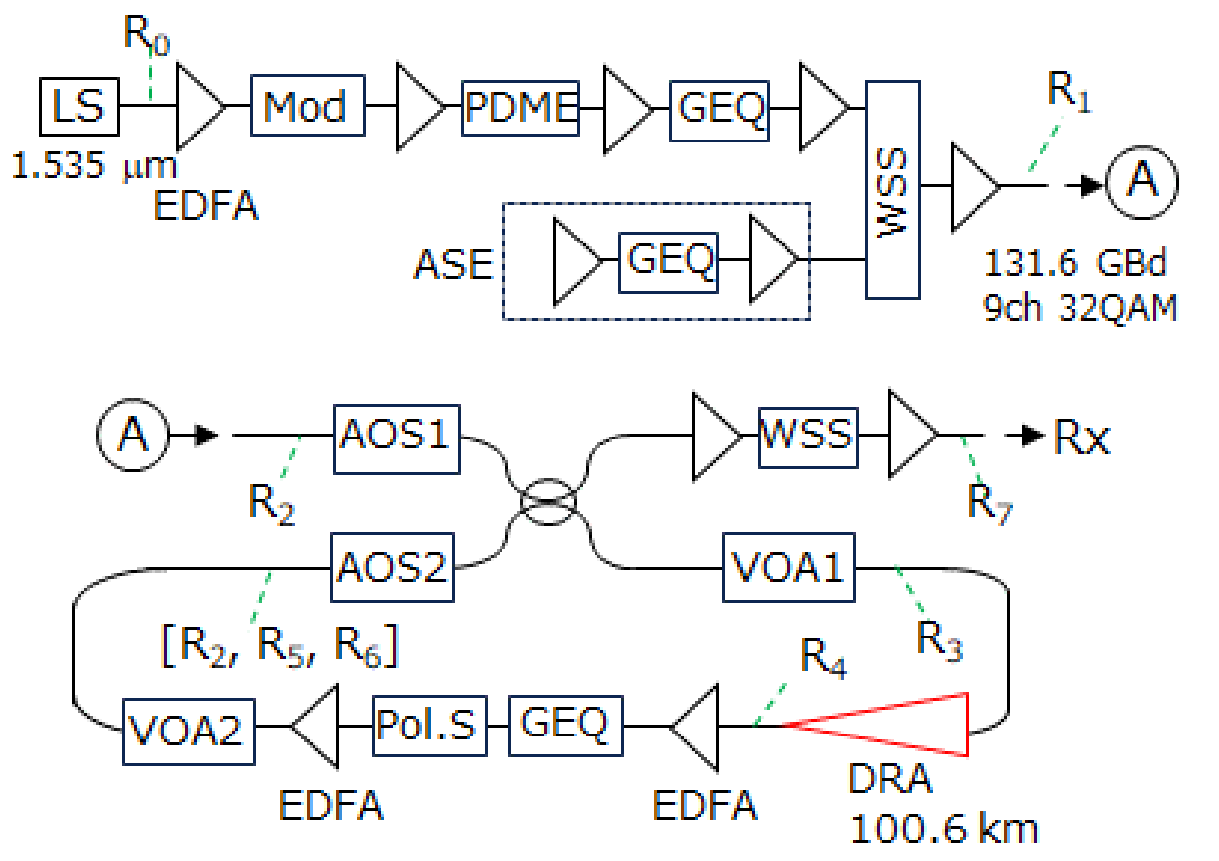

Fig. 10. Experimental setup for data transmission. LS: light source, Mod: optical IQ-modulator, PDME: polarization-division-multiplexing emulator, GEQ: gain equalizer, WSS: wavelength selective switch, AOS: acousto-optic switch, Pol.S: polarization scrambler.

Note that the NFs discussed in this section take into account the optical noise caused by ASE, but they do not take into account the optical noise caused by mutual interference between the pump light and the signal light, which is noticeable in Fwd pumping [8], [13], [14], [15], [16]. Therefore, although evaluation based on NF is important, the final determination of superiority requires measurement of the Q-factor of the transmitted signal, as will be shown in the next section.

## VI. Experimental Setup for Data Transmission Using DRA and EDFA

Fig. 10 shows the experimental setup used for data transmission, including the recirculating loop consisting of a DRA and EDFAs. Reference points $R_0$ - $R_7$ in Fig. 8 are also shown. Note that, $R_2$ in Fig. 8 appears in two places in Fig. 10. One of them is outside the loop, and the other is inside the loop. The optical signal before propagating through the DRA passes through $R_2$ outside the loop, and the optical signal after propagating through the DRA passes through $R_2$ inside the loop until it reaches the last lap. $R_2$ inside the loop in Fig. 10 is equivalent to $R_5$ and $R_6$ in Fig. 8.

The IQ-modulator (Mod) modulated the continuous wave light at 1535-nm output from the light source (LS) and generated the 131.6-GBd PS 32-QAM signal used for the measurement channel. Details on the RF circuits used to drive the IQ-modulator are omitted in Figs. 8 and 10 because they are not the focus of this paper. Before and after the IQ-modulator, polarization-maintaining EDFAs were used to achieve a high OSNR at reference point $R_7$. The modulated signal was sent to the polarization-division-multiplexing emulator (PDME) and multiplexed with itself after passing the delay line inside the PDME. To emulate the 9-ch, 150-GHz-spaced WDM signal, ASE noise from 1530 nm to 1540 nm was generated by other EDFAs. After shaping the spectrum using the gain equalizers (GEQs), the measurement channel and ASE were multiplexed by a wavelength selective switch (WSS). The input and circulation of the signal light were controlled by acousto-optic switches (AOSs) 1 and 2. The structure of the DRA in the recirculating loop was already shown in Fig. 1. VOA1 and 2 were used to control the signal launch power and adjust the loop gain $G_{25}$ to 1 (0 dB). Because of the output power limitation of the EDFAs in the recirculating loop, the highest signal launch power at $R_3$ was +4.0 dBm/ch. The GEQ and polarization scrambler (Pol.S) in the recirculating loop were used to suppress the gain tilt and to cancel the polarization dependent loss, as mentioned before. The transmitted signal output from the recirculating loop was sent to the WSS, and the measurement channel was demultiplexed. After reference point $R_7$, the measurement channel was demodulated using a coherent detection technique, and the Q-factor was calculated using an off-line computer.

## VII. Results and Discussion

Before showing the Q-factor of the transmitted signal, we show the upper limit of the Q-factor that can be achieved with the experimental setup shown in Fig. 10. When AOS1 was “on” and AOS2 was “off”, OSNR at $R_7$ was larger than 40 dB / 0.1 nm. It was high enough to saturate the Q-factor [25]. In this case, the optical signal had no transmission penalty, and the Q-factor was determined by the background noise in the electrical circuit or imperfections in the optical devices. In this case, the measured Q-factor was 11.6 dB. The Q-factor of the transmitted signal cannot exceed this limit because ASE and nonlinear penalties accumulate during multi-span transmission.

Fig. 11 shows the measured Q-factor as a function of the signal launch power and transmission span (number of laps). The limit of the Q-factor is shown as a bold line and labeled as “Upper limit of Q.” In this measurement, the ratio of Fwd pumping to Bwd pumping was changed, though the total on/off gain of the DRA was fixed to 6.4 dB. The cross symbols show the results without using Fwd pumping. The power of Bwd pumping was 408 mW. The circle symbols show the results when the powers of Fwd and Bwd pump light were 276 mW and 151 mW, respectively, in which the on/off gains induced by Fwd and Bwd pumping were 4.0 and 2.4 dB, respectively. For Fwd pumping, the proposed pump unit shown in Fig. 1 was used (filled circle). For comparison, the results when a conventional pump unit was used for Fwd pumping are also shown by the open circles.

Fig. 11 shows that Q-factors after 1-span (100.6-km) transmission were almost saturated to the upper limit of the Q-factor. In this study, the Q-factors were prone to saturate due to the shorter transmission span compared with our previous work (span was 125 km in [25]). The measured Q-factors after transmission over 7 spans (704.2 km) and 23 spans (2313.8 km) were not saturated because of the accumulated ASE and nonlinear penalties. In these cases, the measured Q-factors depended on the signal launch power.

In this paper, “optimum signal launch power” is defined as the power at which the maximum Q-factor can be achieved after transmission. When the signal launch power is lower than the optimum value, the Q-factor is mainly determined by the OSNR [25]. On the other hand, when the signal launch power is higher than the optimum value, nonlinear optical effects become dominant. So then, when the signal launch power is at an optimum value, the Q-factor is influenced by both OSNR and nonlinear optical effects. Fig. 11 shows that when the on/off gains induced by Fwd and Bwd pumping were 4.0 and 2.4 dB respectively and the proposed Fwd pump unit was used (filled circle), the optimum signal launch power was 1.7 dBm/ch. In this case, the maximum Q-factor after 23-span transmission was 8.6 dB (denoted as “A”). When only Bwd pumping was used and the on/off gain was 6.4 dB (cross), the optimum signal launch power was 3.7 dBm/ch. In this case, the maximum Q-factor after 23-span transmission was 8.5 dB (denoted as “B”). When the on/off gains induced by Fwd and Bwd pumping were 4.0 and 2.4 dB, respectively, and the conventional pump unit was used for Fwd pumping (open circle), the optimum signal launch power was 2.7 dBm/ch. In this case, the maximum Q-factor after 23-span transmission was 8.4 dB (denoted as “C”). The optimum signal launch power of “A” was lower than that of “B.” This result confirms that $NF_{07}$ was reduced by introducing the proper Fwd pumping. Moreover, the Q-factor of “A” was slightly higher than that of “B.” This means that the nonlinear penalty (for example, cross phase modulation between adjacent signal channels) was reduced due to the lower signal power. The optimum signal launch power of “C” was higher than that of “A,” though their distributed gains were the same. This means that the noise transfer from the Fwd pump light was greater than “A,” so a higher OSNR was required to achieve the maximum Q-factor. Nevertheless, the Q-factor of “C” was lower than that of “B.” It was reconfirmed that the Q-factor was degraded when Fwd pumping was performed with an improper pump light [4], [24], [25].

In Fig. 11, because the total on/off gain was constant at 6.4 dB, the improvement in the highest Q-factor was not so significant. However, importantly, the optimum signal launch power could be reduced by 2 dB (from 3.7 to 1.7 dBm/ch) by introducing the

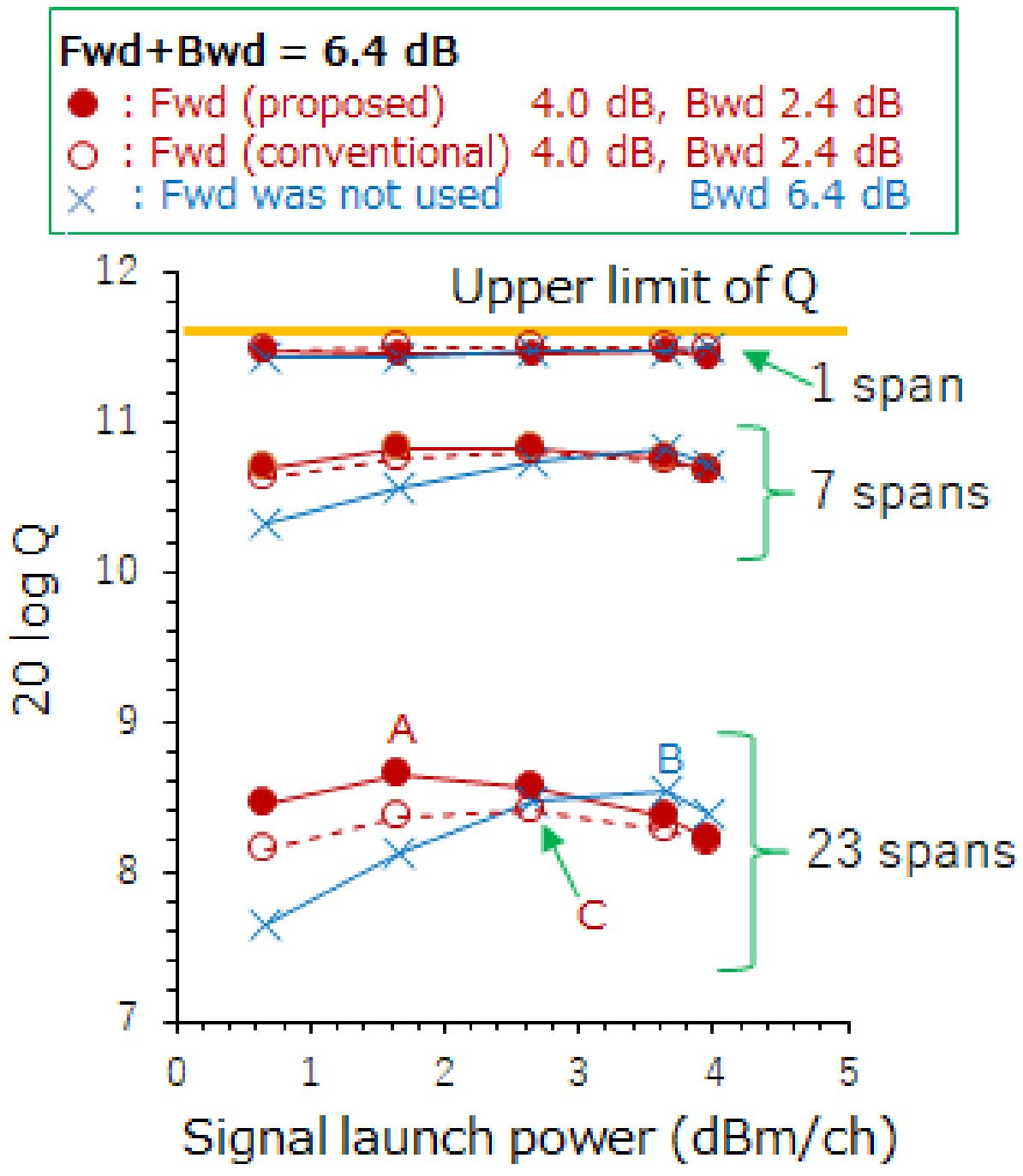

Fig. 11. Measured Q-factor as function of signal launch power and transmission span. Total on/off gain of the DRA was set to 6.4 dB. Circle: on/off gain induced by Fwd pumping and Bwd pumping were 4.0 dB and 2.4 dB, respectively. Cross: Bwd pumping only. Filled circle: proposed Fwd pump unit was used. Open circle: conventional pump unit was used for Fwd pumping.

Fwd pumping. This property can be a great advantage in a transmission system where the power of the pump light and signal light is limited. As mentioned before, due to the limitations of the EDFA, the maximum signal launch power was +4.0 dBm/ch in this experiment. When the number of channels in the WDM signal increases, the maximum signal launch power per channel becomes even more limited. Another limiting factor is the maximum power of the pump light. To amplify multiple wavelength bands with a DRA, pump light of multiple wavelengths is required. However, in commercial transmission lines, a safety limit is required on the total pump power launched at the end of the fiber to prevent accidents caused by heat [29], [30]. When the band of the wavelength becomes wider, the number of pump wavelengths required increases, but the total power of the pump light is limited, so the Raman gain received by each wavelength channel decreases.

Fig. 12 shows the measured Q-factor after 23 spans as a function of signal launch power, when it was assumed that the power of the Bwd pump light dedicated to C-band amplification was limited to about 300 mW and the on/off gain induced by the Bwd pump was limited to 5.0 dB. The cross symbols show the results without using Fwd pumping. The triangle symbols show the results when the power of the Fwd pump light was 134 mW and the on/off gain induced by Fwd pumping was 2.0 dB. The filled circle symbols show the results when the power of the Fwd pump was 276 mW (lower than 300 mW) and the on/off gain induced by Fwd pumping was 4.0 dB. For comparison, the results when a conventional pump unit was used for Fwd pumping are also shown by the open circles. When no Fwd pumping was used, the maximum Q-factor could not be achieved because the optimum signal launch power exceeded the power limit of the signal launch power, 4.0 dBm/ch in this experimental setup (see "E" in Fig. 12). However, with the proposed Fwd pump unit, the maximum Q-factor increased, and the optimum fiber launch power decreased, as the power of the Fwd pump light increased. When the on/off gain caused by Fwd pumping was 4.0 dB, the improvement in the maximum Q-factor was 0.4 dB, and the signal launch power reduced by more than 3.4 dB ("D" in Fig. 12).

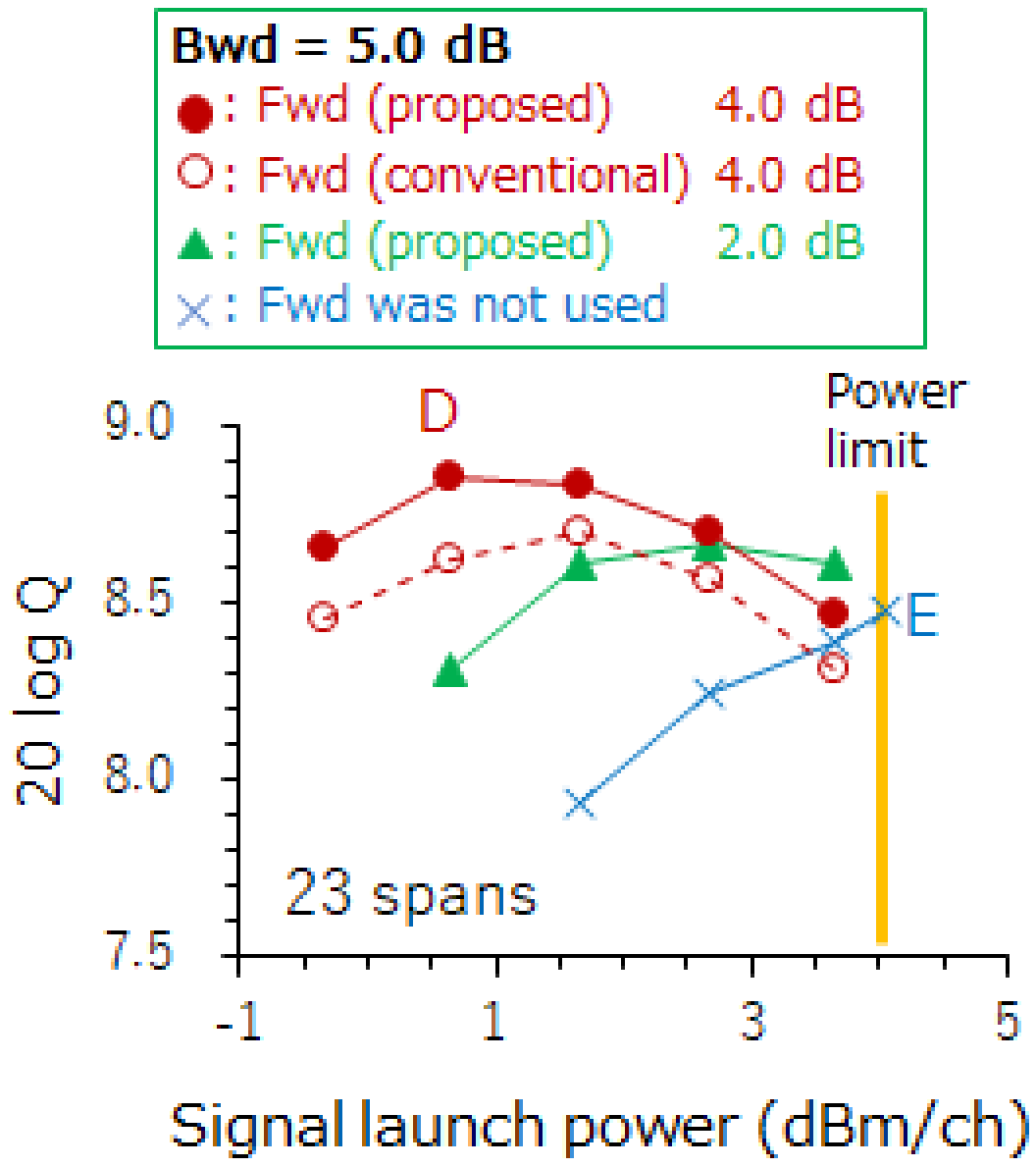

Fig. 12. Measured Q-factor as function of the signal launch power. On/off gain of DRA induced by Bwd pumping was set to 5.0 dB. Circle: on/off gain induced by Fwd pumping was 4.0 dB. Triangle: on/off gain induced by Fwd pumping was 2.0 dB. Cross: Bwd pumping only. Filled circle: proposed Fwd pump unit was used. Open circle: conventional pump unit was used for Fwd pumping.

Finally, we show the measured Q-factor as a function of the OSNR after 1, 7, 9, and 23-span transmissions, in Fig. 13. The on/off gain of the DRA and signal launched power were the same as "D" and "E" shown in Fig. 12. Fig. 13 also shows the Q-factor measured with the back-to-back (B2B) configuration, i.e., 0-span transmission (dot). In the B2B configuration, AOS1 was "on", AOS2 was "off" (see Fig. 10), and the OSNR was controlled by an additional VOA not shown in Fig. 10. When compared at the same OSNR, the Q-factor after transmission was lower than B2B. This deterioration was due to nonlinear penalty. Fig. 13 shows that the nonlinear penalty increased as the number of spans increased. Since the rate of increase was almost independent of the pumping direction, this nonlinear penalty was considered to be mainly due to nonlinear optical effects between WDM signals, rather than the interaction between the signal and pump light that strongly depends on the pumping direction. After 23-span transmission, the measured OSNR when using Fwd pumping together was improved by 0.8 dB compared with using only the Bwd pumping, even though the signal launch power was 3.4 dB lower.

These results indicate that when the proposed low-noise Fwd pumping technique and Bwd pumping were used together in a suitable ratio, the advantages of Fwd pumping outweighed its disadvantages. However, when Fwd pumping was performed with a conventional pump source, the results show that the disadvantages of Fwd pumping outweighed its advantages.

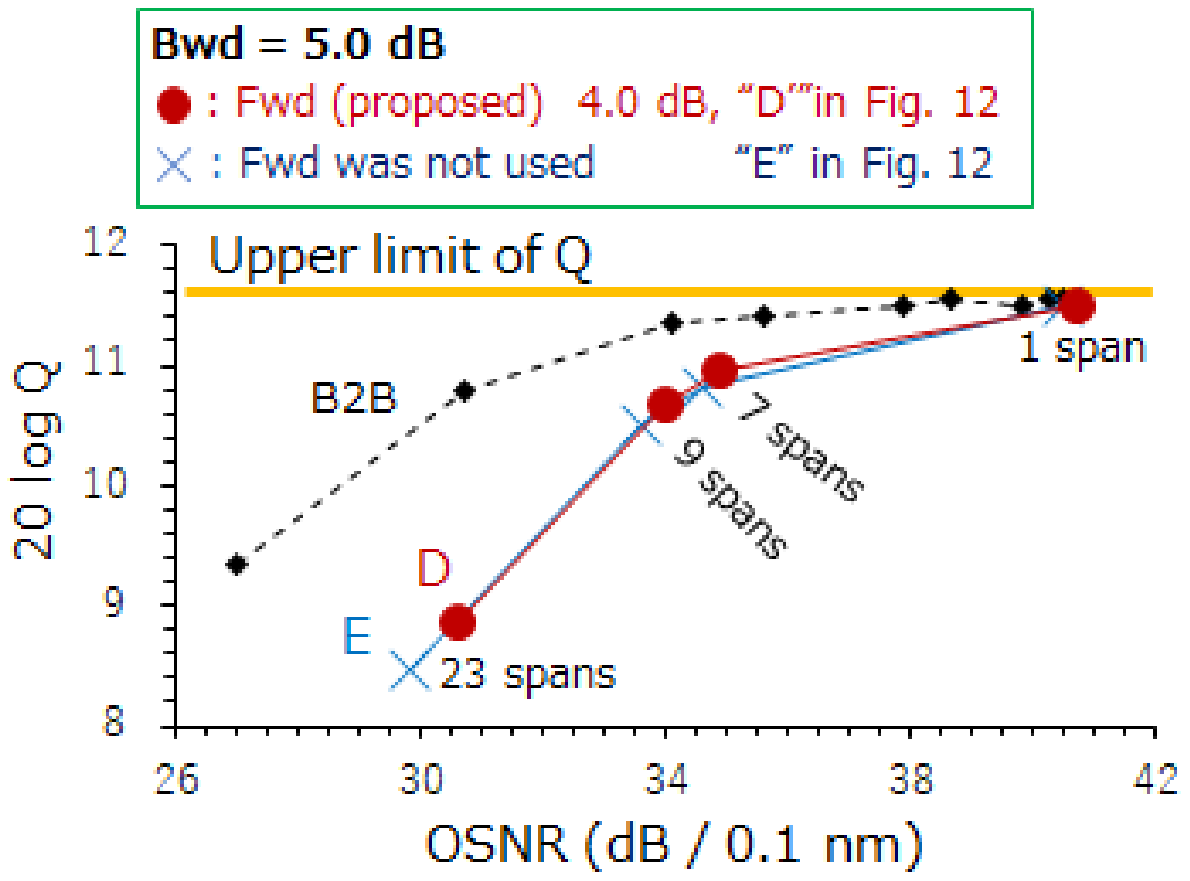

Fig. 13. Measured Q-factor as function of OSNR and transmission span. On/off gain of the DRA and signal launched power were same as "D" and "E" shown in Fig. 12. Circle: on/off gain induced by Fwd and Bwd pumping were 4.0 dB and 5.0 dB, respectively. The signal launch power was 0.7 dBm/ch. Cross: on/off gain induced by Fwd and Bwd pumping were 0.0 dB and 5.0 dB respectively. The signal launch power was 4.0 dBm/ch. Dot: B2B.

Note that, in this study, the Q-factor when using only Fwd pumping in the DRA was not measured. Our previous work suggests that using bi-directional pumping produces better results than using Fwd pumping only [25]. Optimization of the ratio of Fwd pumping and Bwd pumping is for future investigation.

## VIII. CONCLUSION

We constructed a DRA using 100.6-km of G.654.E fiber and our proposed low-noise Fwd pumping technique to investigate its characteristics. When compared under an on/off gain of 6.4 dB, the measured NF with Fwd pumping only was 5.2 dB higher than with the Bwd pumping only.

We measured the Q-factor of the 9-ch WDM signal (131.6 GBd, PDM-PS 32-QAM) after a multi-span transmission line consisting of this DRA and EDFAs while changing the ratio of the Fwd pump power and Bwd pump power in the DRA. After 23-span transmission (2313.8 km), a higher Q-factor (improvement was 0.4 dB) could be achieved with a lower signal launch power (improvement was more than 3.4 dB) by increasing the Fwd pump power. We also showed that when a conventional pump light source was used instead of the proposed Fwd pumping technique, Fwd pumping became a degradation factor.

**Hiroto Kawakami** (Member, IEEE) received the M.S. degrees in physics from Hokkaido University, Sapporo, Japan in 1991.

In 1991, he joined the NTT Transmission Laboratories, Yokosuka, Japan, where he is involved in research and development on high-speed optical communications systems. His research interests include nonlinear effects in optical fiber and device control in transmission systems.

Mr. Kawakami is a member of the Information and Communication Engineers (IEICE) of Japan.

**Kohei Saito** received the M.S. degree in electronics and applied physics from Tokyo Institute of Technology, Tokyo, Japan, in 2013.

In 2013, he joined the Nippon Telegraph and Telephone (NTT) Network Service Systems Laboratories, where he got engaged in research on optical communication systems. He is a Member of the Institute of Electronics, Information and Communication Engineers of Japan. Since 2020, he has been with NTT Network Innovation Laboratories, where he is engaged in research on optical transmission technologies for high-capacity optical transport networks.

He was the recipient of the Young Researcher's Award from IEICE in 2019.

**Akira Masuda** was born in Kanagawa, Japan. He received the B.E. and M.E. degrees in electrical engineering from the Tokyo University of Science, Tokyo, Japan, in 2011 and 2013, respectively. In 2013, he joined NTT Network Innovation Laboratories, NTT Corporation, Yokosuka, Japan, where he was engaged in research on optical technologies for high-speed short-reach transmission. From 2019 to 2021, he joined NTT Communications Corporation, where he was engaged in the development of beyond 100-Gbit/s/ch optical WDM transmission systems. Since 2021, he has been with NTT Network Innovation Laboratories, NTT Corporation, where he is engaged in research on optical transmission technologies for high-capacity optical transport networks. He was the recipient of the Young Researcher's Award from IEICE in 2019.

**Shuto Yamamoto** (Member, IEEE) was born in Tokyo, Japan. He received B.S. and M.S. degrees in physics and a Ph.D. degree in electronic engineering from the Tokyo Institute of Technology, Tokyo, Japan, in 2003, 2005, and 2018, respectively.

He is currently a Senior Research Engineer with Network Innovation Laboratories, Nippon Telegraph and Telephone (NTT) Corporation, Yokosuka, Japan. His research interests include photonic transport network systems.

He is a Member of the IEEE Photonics Society and a Member of the Institute of Electronics, Information and Communication Engineering (IEICE) of Japan. He was a member of the Technical Program Committee for the Optical Fiber Communication Conference (OFC) from 2018 to 2021. He received a Best Paper Award from the 16th Optoelectronics and Communications Conference (OECC), a Young Researcher's Award from IEICE, and an Achievement Award from IEICE in 2011, 2012, and 2016, respectively. He also received an NTT Presidential Award and TELECOM System Technology Award from the Telecommunications Advancement Foundation in 2014 and 2019, respectively.

**Etsushi Yamazaki** (Member, IEEE) received the B.E. and M.E. degrees in electronic engineering from The University of Tokyo, Tokyo, Japan, in 2000 and 2002, respectively.

He is currently with Nippon Telegraph and Telephone Corporation Network Innovation Laboratories, Yokosuka, Japan, where he has been engaged in research and development for optical transmission system, and photonic network. From 2018 to 2019, he was a Visiting Researcher with Stanford University, Stanford, CA, USA.

He is a Member of the Institute of Electronics, Information and Communication Engineers of Japan. In 2012, he was the recipient of the Sakurai-Kenjiro Memorial Award from Optoelectronics Industry and Technology Development Association of Japan.